\documentclass[aps,pra,amsmath,amssymb,showpacs,twocolumn]{revtex4-1}

\usepackage[T1]{fontenc}
\usepackage{amssymb,amsmath,braket}
\usepackage{amsfonts}
\usepackage{graphicx}
\usepackage{color}
\usepackage{epstopdf}
\usepackage{hyperref}

\newcommand{\pw}{{w}}
\newcommand{\Pclas}{{P_{\mathrm{cl}}(\alpha)}}
\newcommand{\updelta}{{\pw_{\mathrm{max}}^{\delta,\mathcal{NC}}}}
\newcommand{\updeltaSC}{{\pw_{\mathrm{max}}^{\delta,\mathcal{SC}}}}
\newcommand{\updeltasqueezed}{{\pw_{\mathrm{max, \, squ}}^{\delta,\mathcal{NC}}}}
\newcommand{\nscgaussvac}{{\pw_\mathrm{max, \,vac}^{\mathrm{G},\, \mathcal{NSC}}}}
\newcommand{\upncgauss}{{\pw_\mathrm{max}^{\text{G},\,\mathcal{NC}}}}

\def\ket#1{|#1\rangle}
\def\bra#1{\langle#1|}

\renewcommand{\leq}{\leqslant}
\renewcommand{\geq}{\geqslant}

\renewcommand{\ge}{\geqslant}

\newenvironment{proof}[1][Proof]{\noindent\textbf{#1.} }{\ \rule{0.5em}{0.5em}}

\begin{document}

\title{Nonclassical states of light with a smooth $P$ function}
\author{Fran\c{c}ois Damanet$^{1,2}$, Jonas K\"ubler$^{3}$, John
  Martin$^{2}$, and Daniel Braun$^{3}$}
\affiliation{$^{1}$ Department of Physics and SUPA, University of Strathclyde, Glasgow G4 0NG, United Kingdom.\\
$^{2}$ Institut de Physique Nucl\'{e}aire, Atomique et de Spectroscopie, CESAM,
Universit\'{e} de Li\`{e}ge, B\^{a}timent B15, B - 4000 Li\`{e}ge, Belgium.
\\
$^{3}$ Institut f\"ur theoretische Physik, Universit\"at T\"{u}bingen,
72076 T\"ubingen, Germany.
}
\begin{abstract}
There is a common understanding in quantum optics that nonclassical
states of light are states that do not have a positive semidefinite
and sufficiently regular
Glauber-Sudarshan $P$ function.  Almost all known nonclassical states
have $P$ functions that are highly irregular which makes working with
them difficult and direct experimental reconstruction impossible. Here
we introduce classes of nonclassical states with regular,
non-positive-definite $P$ functions.  They are constructed by
``puncturing'' regular smooth positive $P$ functions with negative Dirac-delta peaks, or other sufficiently narrow smooth
negative functions. We determine the parameter ranges for which such
punctures are possible without losing the positivity of the state, the regimes yielding antibunching of light, and the expressions of the Wigner functions for all investigated punctured states. Finally, we propose some possible experimental realizations of such states.

\end{abstract}

\maketitle
\section{Introduction}

Creating and characterizing nonclassical states of light are two tasks of substantial interest in a wide range of applications~\cite{Str17, Pat17}, such as sub-shot-noise measurements~\cite{Xia87, Sha09, Ono17}, high resolution imaging~\cite{Bot00, Str01, Gen16, Tay14}, light source and detector calibration~\cite{Zel69,Bur70,Kly80}, quantum cryptography \cite{Bra00, Zhe13}, or quantum information processing~\cite{Bar03}.

In quantum optics, there is wide-spread agreement that the
most classical pure states of light are coherent states
$\ket{\alpha}$.  These are eigenstates with complex eigenvalue
$\alpha$ of the annihilation operator $a$ of the 
mode of the light-field under consideration, 
$a\ket{\alpha}=\alpha\ket{\alpha}$, and are
characterized by the minimal uncertainty product of the quadratures
$X=(a+a^\dagger)/\sqrt{2}$ and $Y=(a-a^\dagger)/(i\sqrt{2})$.
In the case of a mechanical harmonic oscillator the quadratures
correspond simply to phase space coordinates $x$ and $p$ in suitable
units. Hence, a coherent state resembles most closely the notion of a classical
phase space point located at the mean position of the quadratures.  The label
$\alpha\in\mathbb C$ determines these mean values as
$\braket{\alpha|X|\alpha}{}=\sqrt{2}\,\mathrm{Re} (\alpha)$ and
$\braket{\alpha|Y|\alpha}{}=\sqrt{2}\,\mathrm{Im} (\alpha)$. 
Coherent states of light, which can be produced by lasers operated far above the threshold~\cite{Chi08}, retain the property of minimal uncertainty $\Delta
X\Delta Y=1/2$, where $\Delta X = \sqrt{\langle X^2\rangle - \langle X \rangle^2}$ is the standard deviation,
under the dynamics of the harmonic oscillator underlying each mode of the
electromagnetic field. Mixing classically coherent states (i.e.~choosing randomly such states
according to a classical probability distribution) should not increase
their quantumness.  Hence, (possibly mixed) quantum states given by a
density operator $\rho$ as
\begin{equation}\label{eq:rho_all}
\rho = \int P(\alpha) \ket{\alpha}\bra{\alpha}\,d^2 \alpha 
\end{equation}
with a positive $P$ function are considered ``classical states'' in
quantum optics.  While it may appear that \eqref{eq:rho_all}
describes only states that are ``diagonal'' in a coherent state basis,
it was shown by Sudarshan \cite{Sud63} that in fact all possible
quantum states 
$\rho$ of the light-field (quantum or classical) can 
be represented by \eqref{eq:rho_all}.  This is possible
due to the fact that coherent states are overcomplete. The price to
pay, however, is that in general $P(\alpha)$ is not a smooth function,
but a functional.  Obviously, indeed, even a coherent state
$\ket{\alpha_0}$ is already represented by a functional, namely a
Dirac-delta peak, $P(\alpha)=\delta(\alpha-\alpha_0)$. Such a level of
singularity  represents the ``boundary of acceptable irregular
behavior'' of the 
$P$ function for a classical state. Much worse singularities arise for
example from simple Fock states $\ket{n}$, i.e.~energy eigenstates of the mode,
$a^\dagger a\ket{n}=n\ket{n}$, for which the $P$ function is given by the $n$-th
derivative of a delta-function~\cite{Aga12}. \\

It would be
desirable that one could directly reconstruct the
$P$ function and in this way show the nonclassicality of states that
are genuinely quantum.  However, in the case of highly non-regular
$P$ functions this is impossible.  One way out is to apply filters
that do not enhance the negativity  
of the characteristic function, i.e.~the Fourier
transform of $P(\alpha)$~\cite{Kie10,Kie11,Agu13,Kuh14}. 

Also for the ease of theoretical work one would like to have
classes of $P$ functions that are regular  and smooth, but
represent 
genuine quantumness due to the fact that they become negative~\cite{Kly96}. One
class of such states are single-photon added thermal states~\cite{Kie11},
i.e.~states of the form $\rho=\mathcal N a^\dagger
\rho_T a $, where $\rho_T$ represents a thermal state and
$\mathcal N$ is a normalization factor. The $P$ function of this state
is given
by the product of a linear function of $|\alpha|^2$ and the Gaussian
of the thermal state (see Eq.~(4.36)
in \cite{Aga12}). 
But as Agarwal writes (p.84 in \cite{Aga12}): ``Here we
perhaps 
have the unique case where $P(\alpha)$ 
exists but is negative. Most known cases of nonclassical $P$ involve
$P$ functions which do not even exist.`` \\

In this paper, we introduce novel whole classes of such genuinely-quantum states with nevertheless smooth $P$ functions (i.e.~exhibiting at most singularities in the form of Dirac $\delta-$functions). The recipe for doing so is very simple: Start
with a classical quantum state with a smooth positive $P$ function, and
then superpose some narrow, negative peaks.  Altogether, one must
guarantee, of course, that the density matrix $\rho$ remains a semidefinite positive operator.  Finding parameter regimes where this is the case
constitutes the main technical difficulty for constructing such
states. Obviously, this general recipe allows for an infinity of
possibilities, starting from the choice of the smooth original $P$ function, over the number and form of the
negative peaks superposed, to their amplitudes and positions.
We start with the simplest situation of the rotationally symmetric
Gaussian $P$ function centered at $|\alpha|=0$ of a 
thermal state, ``punctured''  with a single negative delta-peak whose
position and amplitude we can vary. Only the radial position is relevant
here, and we have herewith already three parameters, exemplifying the wide range of tunability of punctured states.

The paper is organized as follows. In Section II, we introduce the ``punctured states'', their general properties, the conditions for positive definiteness of their density matrix representation and the conditions yielding nonclassical states. In Section III, we investigate the particular simple case of a thermal state punctured by a single delta-peak. In Section IV, we study the case of a squeezed thermal state punctured by a delta-peak. In Section V, we generalize our results by considering a thermal state punctured by a Gaussian peak of finite width. In each Section, we provide the expression of the Wigner function, calculate the second-order correlation function and identify the regimes of parameters yielding antibunching of light, i.e.\ sub-Poissonian statistics of the photon counting. In Section VI, we discuss some possibilities of experimentally creating some of these states. Finally, in Section VII, we give a conclusion of our work.

\section{Definition, general properties, positivity and nonclassicality conditions}

\subsection{General punctured states}
Let $\rho_{\mathrm{cl}}$ represent a classical state, that is a state as defined in \eqref{eq:rho_all} with a smooth non-negative $P$ function $P_{\mathrm{cl}}(\alpha) \geq 0$. We define a ``punctured $\rho_{\mathrm{cl}}$ state'' as a state~(\ref{eq:rho_all}) with a $P$ function of the form
\begin{equation}\label{eq:rho_punctured}
P(\alpha) =
\mathcal{N}\left[\Pclas -\sum_{i=1}^N \pw_{i}\,\pi_i(\alpha-
  \alpha_{i})\right],
\end{equation}
where $\mathcal{N}$ is a normalization constant, the $\pw_{i}$ are positive coefficients representing the weights of the different punctures $i$ ($i = 1, \dotsc, N$), and $\pi_i(\alpha)$ are positive functions on the complex plane that determine their shape. We take them as normalized according to
\begin{equation}
  \label{eq:normpi}
  \int \pi_i(\alpha)\,d^2\alpha =1\, \quad\quad \forall i = 1, \dotsc, N.
\end{equation}
 
If the weights are chosen such that the $P$ function \eqref{eq:rho_punctured} becomes negative for some $\alpha$ and that the resulting state $\rho$ is a positive semidefinite operator with unit trace, it follows immediately that the state is nonclassical.
Furthermore, the $P$ function remains
``well-behaved'' (i.e.~sufficiently regular), if $\pi_i(\alpha)$ are
well-behaved, as we assumed $\Pclas$ smooth. In agreement with the
common understanding that positive $P$ functions
corresponding to simple Dirac-$\delta$ peaks are still accepted in the
class of ``well-behaved'' functions because they represent coherent
states, we consider in the following sections Dirac $\delta(\alpha)$ functions as worst singularities for the punctures $\pi_i(\alpha)$. This is the
also the simplest case, as it does not contain
any further free parameter.

\subsection{Conditions for (semi)positive definiteness}
For a given
choice of the functions $\pi_i(\alpha)$ for the punctures, 
positivity of the state will depend on the position of the
$\alpha_i\in\mathbb C$ and the 
weights $\pw_i$. Our task is hence to find the allowed
parameter ranges for $\{\pw_i,\alpha_i\}$ such that $\rho\ge 0$, i.e.\ $\rho$ is a positive semidefinite operator. The
normalization factor  $\mathcal{N}$ ensures
that $\int P(\alpha)d^2\alpha=1$, i.e.
\begin{equation}\label{normcond}
\mathcal{N}=\left(1-\sum_{i=1}^N \pw_i \right)^{-1}\,.
\end{equation}
Since $\mathcal{N}$ does not influence the positivity, we will often
skip it.

It is clear that in general the condition for
any single parameter
will depend on the values of all the others. This can be seen
immediately from the general necessary and sufficient condition of positive semidefiniteness, $\langle \psi |\rho|\psi\rangle \geq 0 \;\forall\: |\psi\rangle\in\mathcal{H}\equiv L^2(\mathbb{C})$, which for the state~(\ref{eq:rho_all}) with a $P$ function of the form (\ref{eq:rho_punctured}) reads 
\begin{equation}\label{poscond}
\int \Big(\Pclas-\sum_{i} \pw_i \pi_i(\alpha-
  \alpha_{i})\Big)\, |\langle \psi |\alpha\rangle|^2 d^2 \alpha \geqslant 0  \;\forall \: |\psi\rangle\in\mathcal{H}\,.
\end{equation}
To go beyond this general criterion, we first review briefly a few
necessary and sufficient conditions for positivity of a linear operator.

{\em Necessary and sufficient condition ($\mathcal{NSC}$).}
 A necessary and sufficient condition ($\mathcal{NSC}$) for a hermitian matrix $\rho$ to be semi-positive definite ($\rho\ge 0$) is that all its eigenvalues are
non-negative. While the eigenvalues of an infinite dimensional matrix
are generally not readily accessible, the condition nevertheless allows for a numerical approach. Indeed, the density operator~(\ref{eq:rho_all}) can be expressed in the Fock state basis $\{\ket{n}:n\in \mathbb{N}_0\}$ using the expansion of coherent states
\begin{equation}\label{alphaasFock}
\ket{\alpha} = e^{-\frac{|\alpha|^2}{2}} \sum_{n = 0}^\infty \frac{\alpha^n}{\sqrt{n!}} \ket{n}.
\end{equation}
Then, by introducing a cut-off dimension in Hilbert space of sufficiently high excitation $n_\text{max}$, a finite dimensional matrix representation can be obtained and one can check numerically that when increasing the cut-off, the lowest eigenvalue converges to a
value that is sufficiently far from zero for judging the positivity of
the state.

{\em Necessary condition ($\mathcal{NC}$).}
Since $\rho\ge 0 \Leftrightarrow \langle \psi |\rho|\psi\rangle \geq 0$ for all states $|\psi\rangle$, a
necessary condition ($\mathcal{NC}$)  for $\rho\ge 0$ is that all
diagonal elements $\rho_{\phi\phi} \equiv\braket{\phi|\rho|\phi}$ of  
the density matrix are non-negative for a given class of states $\ket{\phi} \in \mathcal{K} \subset \mathcal{H}$ where $\mathcal{K}$ is a subset of $\mathcal{H}$, i.e.
\begin{equation}\label{NC}
\rho_{\phi\phi} \equiv\braket{\phi|\rho|\phi} \geqslant 0, \; \forall\, \ket{\phi} \in \mathcal{K} \quad  \Leftarrow \quad  \rho \geqslant 0.
\end{equation}

{\em Sufficient condition ($\mathcal{SC}$).}
A sufficient condition ($\mathcal{SC}$) for positivity can be found
from Gershgorin's Circle 
Theorem. Consider first the case of finite dimensional matrices $A$.  The
theorem then says \cite{Bra46}: ``Every 
  eigenvalue of a complex square matrix $A$ of coefficients
  $\{a_{ij}\}$ lies within at least one of the Gershgorin disks
  $D(a_{ii}, R_{i})$ centred at $a_{ii}$ with radii $R_{i} = \sum_{j
    \neq i} |a_{ij}|$ or $R_{i} 
  = \sum_{j \neq i} |a_{ji}|$.'' An interpretation of this theorem is
that if the off-diagonal elements of $A$ have small norms
compared to its diagonal elements, its eigenvalues cannot be far from
the values of the diagonal elements.  
A matrix is called (strictly) diagonally dominant if $|a_{ii}| \geq
R_{i}$ ($|a_{ii}| >
R_{i}$) for all $i$.
This leads to the following sufficient condition:
a hermitian (strictly) diagonally dominant matrix with real non-negative diagonal entries is positive semidefinite
(definite). Gershgorin's Circle Theorem has been generalized to the case of
infinite dimensional matrices in~\cite{Shi87}, where a proof was given for matrices of the space $L^1$. The proof for the space $L^2$ in which our operators live can be done analogously and leads to the same result, namely that the eigenvalues lie in $\cup_{i = 1}^{\infty}
D(a_{ii}, R_i)$ (just as in the case of a finite space). Hence, by expressing the density matrix in the Fock state basis, we obtain the sufficient condition ($\mathcal{SC}$)
\begin{equation}\label{SC}
\rho_{nn} - R_n \geqslant 0 \quad \forall\, \ket{n} \quad  \Rightarrow \quad  \rho \geqslant 0,
\end{equation}
where $\rho_{nn} \equiv \langle n | \rho | n \rangle$ and $R_{n} = \sum_{m
    \neq n} |\rho_{nm}|$.

\subsection{Conditions for nonclassicality}

There exist various nonclassicality criteria for radiation fields. In this work, we shall consider criteria based on the negativity of the $P$ function, the negativity of the Wigner function and the existence of antibunching~\cite{Str17}.

{\em Negativity of the $P$ function.}
A criterion for a state to be nonclassical is that its $P$ function takes negative values, so that it can no longer be interpreted as a probability distribution in phase space. For example, all physical $\delta$-punctured states are nonclassical according to this criterion. Other punctured states, however, could have a non-negative $P$ function in the whole complex plane, so that general punctured states are not necessarily nonclassical. 

{\em Negativity of the Wigner function.}
Negativity of the Wigner function is a widely used criterion for nonclassicality. This is largely due to the fact that the Wigner function can be directly measured~(see e.g.~\cite{Smi93,Wal96,Ban96,Lut97,Ban99,Ber02}). Since the Wigner function $W(\alpha)$ is the convolution of the $P$ function with the vacuum state, i.e.\
\begin{equation}\label{Wfunc}
W(\alpha) = \frac{2}{\pi} \int P(\beta)\, e^{-2 |\alpha - \beta|^2} d^2\beta,
\end{equation}
the existence of negative values for the Wigner function implies the
existence of negative values for the $P$ function, but the reverse is
not true. Thus, the criterion based on negative values of the Wigner
function detects fewer nonclassical states than the previous
criterion. More specifically, it does not detect states with partly
negative $P$ function but everywhere positive Wigner function. This
explains why the definitions of punctured states based on the
puncturing of $P$ functions rather than Wigner functions is more
appropriate. Moreover, delta-puncturing of a Wigner function would be
forbidden as Wigner functions are always bounded (see e.g.~p.~73 in
\cite{Schleich}). 

{\em Antibunching.}
Another standard criterion of nonclassicality is based on the second order correlation function
\begin{equation}\label{g2def}
g^{(2)} = \frac{\langle a^\dagger a^\dagger a a\rangle}{\langle a^\dagger a \rangle^2},
\end{equation}
where $\langle\,\cdot \, \rangle = \mathrm{Tr}\left( \,\cdot \,\rho \right)$, or, equivalently, on the Mandel $Q_M$ parameter~\cite{Man79}
\begin{equation}
Q_M  =\frac{(\Delta n)^2 - \langle n\rangle}{ \langle n\rangle} = \langle n \rangle \left( g^{(2)}-1\right),
\end{equation}
where $n = a^\dagger a$. A state is said to be nonclassical if $g^{(2)} < 1$ (or, equivalently, $Q_M <0$), which corresponds to antibunching (sub-Poissonian statistics of photon counting that cannot be observed for mixtures of coherent states).
\\
It is interesting to note the connection between classicality in terms
of the $P$ function and in terms of antibunching, pointed out by
Kimble {\em et al.} \cite{Kim77}, and generalized by Vogel to
space- and time-dependent correlations \cite{Vog08}: $g^{(2)}<1$ can exist if the $P$ function is negative somewhere. Conversely, \emph{if the $P$ function of a state is a valid probability density, then there is no antibunching, i.e.~$g^{(2)} \geq 1$.}\\ 
\begin{proof}
This statement can be proved using Jensen's inequality. We use equation (5') in the original paper~\cite{Jen06}. Let $P(\alpha)$ be a valid probability density defined over the complex numbers. Then $P_r(r) = \int_0^{2\pi}  r P(re^{i\phi})\,d\phi$ defines a valid probability density over the real numbers $r\in [0, \infty)$. Using Eq.~(\ref{eq:rho_all}), the second order correlation function~(\ref{g2def}) can be written as
\begin{align}
g^{(2)} = \frac{ \int  |\alpha|^4 P(\alpha)\,d^2\alpha}{\left[\int  |\alpha|^2 P(\alpha)\,d^2\alpha\right]^2} = \frac{ \int_0^\infty  r^4 P_r(r)\,dr}{\left[\int_0^\infty  r^2 P_r(r)\,dr\right]^2}.
\end{align}
Defining the convex functions $f(r)=\varphi(r)= r^2$, we can rewrite the above as
\begin{align}
g^{(2)} = \frac{\int_0^\infty \varphi(f(r))\, P_r(r)\,dr}{\varphi\left(\int_0^\infty f(r)\,P_r(r)\,dr\right)}.
\end{align}
The condition $g^{(2)} \geq 1$ is then equivalent to 
\begin{align}
\int_0^\infty \varphi(f(r))\, P_r(r)\,dr \geq \varphi\left(\int_0^\infty f(r)\,P_r(r)\,dr\right), 
\end{align}
which is true by Jensen's inequality.
\end{proof}\\

As a conclusion, for radiation field states, both the negativity of the Wigner function and antibunching imply that the $P$ function is not a valid probability density, while the reverse is not necessarily true. 

In the following sections, we investigate the cases of various punctured states. In each case, we apply the conditions of positivity of and nonclassicality of $\rho$ to determine the physical nonclassical states. For the sake of simplicity, we restrict ourselves in the remainder of the paper to a single puncture by setting $N=1$ in Eq.~(\ref{eq:rho_punctured}).

\section{Delta-punctured thermal states}

\label{sec:deltap}

We define a \emph{$\delta$-punctured thermal state} as a state determined through~(\ref{eq:rho_all}) and (\ref{eq:rho_punctured}) with
\begin{equation}\label{eq:palpha}
\begin{aligned}
& \Pclas = \frac{e^{-|\alpha|^2/\bar{n}}}{\pi \bar{n}}, \, \\
& \pi(\alpha)=\delta(\alpha),
\end{aligned}
\end{equation}
where $\bar{n} = (e^{\hbar \omega/k_B T} - 1)^{-1}$ is the average number of
thermal excitations at temperature $T$ of the radiation field with frequency
$\omega$.

\subsection{Conditions for (semi)positive definiteness}

{\em Necessary condition.}
A necessary condition for the positivity of $\rho$ can be obtained from the condition~(\ref{NC}) with $\mathcal{K}$ the set of coherent states. It yields the following upper bound for the puncture weight $w_1$:
\begin{equation}
{
\pw_1 \leq \frac{e^{-|\alpha_1|^2/\bar{n}}}{\bar{n}+1} \equiv \updelta.
}
\label{eq:b1}
\end{equation}

\begin{proof} 
Let $\ket{\gamma}$ ($\gamma\in\mathbb{C}$) denote a coherent state. The density matrix element $\rho_{\gamma\gamma} \equiv\bra{\gamma} \rho \ket{\gamma}$ is then given by
\begin{equation} 
\rho_{\gamma\gamma} = \mathcal{N}\left[\frac{1}{\pi \bar{n}} \int e^{ - \frac{|\alpha|^2}{\bar{n}} - |\alpha - \gamma|^2}  d^2\alpha   -  \pw_1 \,e^{-|\alpha_1 - \gamma|^2}\right],
\end{equation}
where we used $\left|\langle \gamma \ket{\alpha}\right|^2 = e^{-|\alpha - \gamma|^2}$. The condition that $\rho_{\gamma\gamma}$ must be non-negative for all $\gamma$ reads
\begin{equation}
\pw_1 \leq  \frac{1}{\pi \bar{n}} \frac{ \int e^{ - \frac{|\alpha|^2}{\bar{n}} - |\alpha - \gamma|^2}  d^2 \alpha}{e^{-|\alpha_1 - \gamma|^2}} \quad \forall \gamma \in \mathbb{C}.
\end{equation}
After integration, this yields the condition
\begin{equation}
\pw_1 \leq  \frac{e^{-\left( \frac{1}{\bar{n}+1} \right) |\gamma|^2 + |\alpha_1 - \gamma|^2}}{\bar{n}+1}  \quad \forall \gamma \in \mathbb{C}.
\end{equation}
The minimum over $\gamma$ of the rhs term is attained for $\gamma = \left(\frac{\bar{n} + 1}{\bar{n}}\right) \alpha_1$ and directly leads to the condition (\ref{eq:b1}).
\end{proof}

\begin{figure}[h!]
  \begin{center}
    \includegraphics[width=0.45\textwidth]{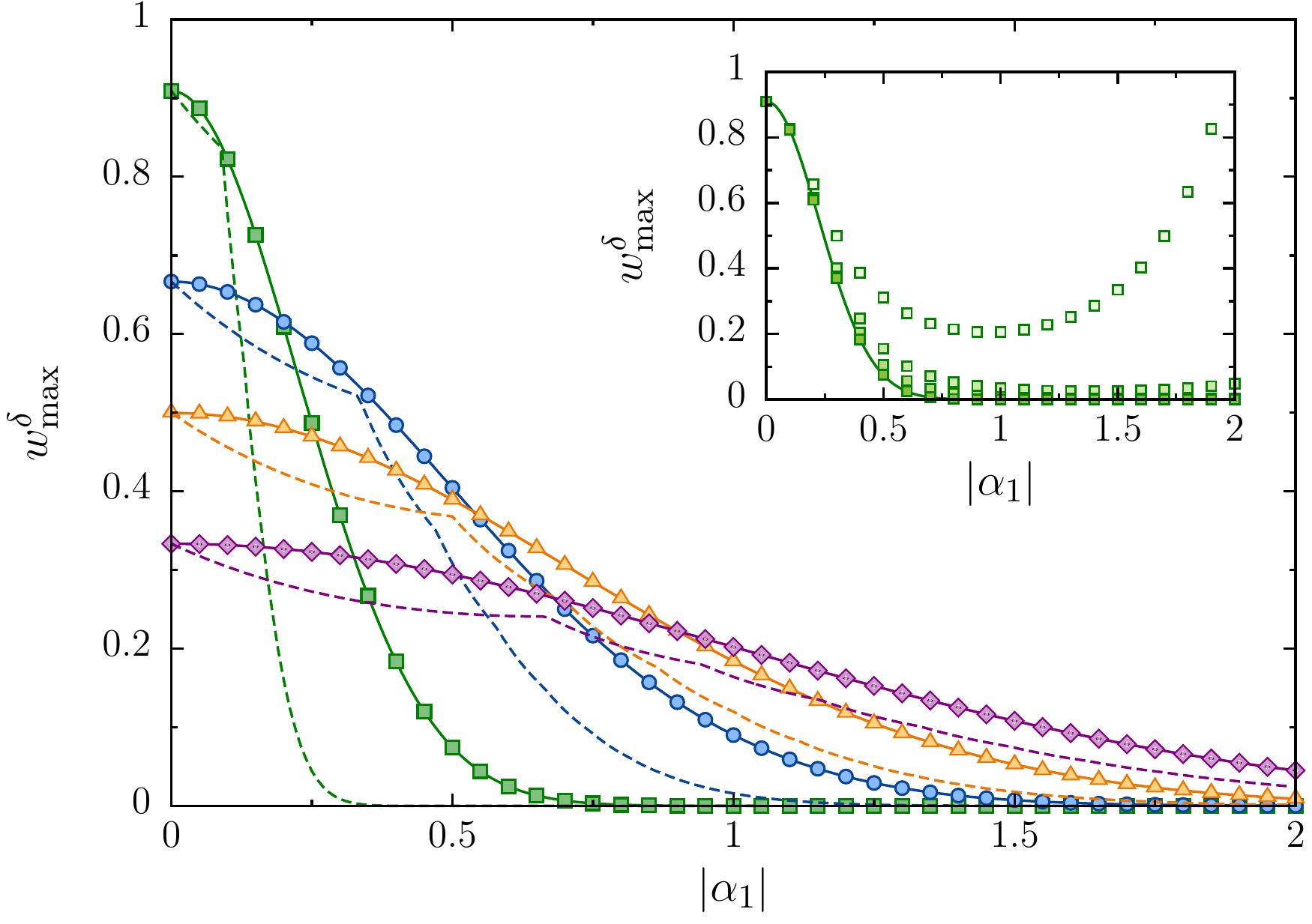}
    \caption{{Plot of the upper bounds $\updelta$ (Eq.~(\ref{eq:b1}), solid curves) and $\updeltaSC$ (Eq.~(\ref{eq:b3}), dashed curves), and of $\pw_{\mathrm{max}}^\delta$, the $\pw_1$ which cancels the smallest eigenvalue of the density matrix, as a function of $|\alpha_1|$ for different $\bar{n}$ (green squares: $\bar{n}=0.1$, blue dots: $\bar{n}=0.5$, orange triangles: $\bar{n}=1$, and purple diamonds: $\bar{n}=2$). The eigenvalues are computed in a Fock space of dimension $n_\mathrm{max} = 15$. Inset: convergence of $\pw_{\mathrm{max}}^\delta$ (green squares) with increasing cut-off dimension $n_\mathrm{max}$ ($n_\mathrm{max}=2,3,4$ and $15$ from top to bottom) for $\bar{n}=0.1$.
        }} 
    \label{fig:tight}
  \end{center}
\end{figure}

{\em Sufficient condition.}
In the Fock state basis, the Gershgorin disks $D(\rho_{nn}, R_{n})$ have center 
$\rho_{nn} \equiv \langle n | \rho | n \rangle $ and radius $R_n$ given by 
\begin{align}
& \rho_{nn} = \frac{\bar{n}^n}{(\bar{n} + 1)^{n+1}}- \pw_1 e^{-|\alpha_1 |^2}\frac{|\alpha_1 |^{2n}}{n!}, \\
& R_n =  \pw_1 e^{-|\alpha_1|^2} \frac{|\alpha_1|^n}{ \sqrt{n!}} \sum_{m = 0 \atop m\neq n}^{+\infty} \frac{ |\alpha_1|^m}{\sqrt{m!}}.
\end{align}
The sufficient condition~(\ref{SC}) then implies the following condition on $\pw_1$,
\begin{equation}
\pw_1 <  \frac{e^{|\alpha_1|^2}}{\bar{n}+1} \frac{1}{f(|\alpha_1|)} \min_{n \in \mathbb{N}} \left\{\left(\frac{\frac{\bar{n}}{\bar{n}+1}}{|\alpha_1|}\right)^n \sqrt{n!} \right\} \equiv \updeltaSC
\label{eq:b3}
\end{equation}
where
\begin{equation}
f(|\alpha_1|) = \sum_{m = 0}^{+\infty} \frac{|\alpha_1|^m}{\sqrt{m!}}\,.
\end{equation}
The bound $\updeltaSC$ depends only on the absolute value of
$\alpha_1$, not on its phase. Also, we have $\updeltaSC = 0$ only if
$\bar{n} = 0$. The largest 
values of $\pw_1$ allowed by the bound lie in the region where
$|\alpha_1|$ and $\bar{n}$ are small, which is shown in
Fig.~\ref{fig:tight}. Since $\updeltaSC > 0$ for all values of 
$\bar{n}>0$ and $\alpha_1$, this is an analytical proof that there always exist physical nonclassical $\delta$-punctured thermal states.

{\em Tightness of $\updelta$.}
\label{tightnessb1}
Computations based on the diagonalization of $\delta$-punctured
thermal states and the application of $\mathcal{NSC}$ suggest that the
bound $\updelta$ given in Eq.~(\ref{eq:b1}) is tight. This claim is
further supported by the observation that the coherent state $\ket{
  \left(\frac{\bar{n} + 1}{\bar{n}}\right) \alpha_1}$ from which the
bound is derived is an eigenstate of $\rho$ with $w_1=\updelta$ with
eigenvalue $0$. In Fig.~\ref{fig:tight},  we display both $\updelta$
and the value of $\pw_1$ that cancels the smallest eigenvalue of
$\rho$. In our computations, the density matrix is expressed in the
truncated Fock state basis $\left\{ \ket{0}, \ket{1}, \dotsc,
  \ket{n_\mathrm{max}}\right\}$ with a sufficiently large cut-off
$n_\mathrm{max}$ in order  ensure stable numerical results. An
analytical proof of the tightness for a puncture at $\alpha_1=0$ will
be provided in the context of Gaussian punctures
(Sec.~\ref{sec:VCGP}). 
Note that a delta-puncture at $\alpha=0$ with
maximum weight $\updelta$ corresponds to the complete removal of the
ground-state of the oscillator (``the vacuum'' in quantum-field theory parlance) from
the state. We call such states ``vacuum-removed states'' for short.

In summary, our results indicate that any $\delta$-punctured thermal
state with $w_1$ satisfying Eq.~(\ref{eq:b1}) corresponds to a proper
physical state. As can be seen in Fig.~\ref{fig:tight}, the value of
the bound of allowed weights $w_1$ decreases as $|\alpha_1|$
increases, showing that a thermal state cannot be punctured
significantly far from its center. This feature is less pronounced as
$\bar{n}$ increases. 

\subsection{Conditions for nonclassicality}

{\em Negativity of the $P$ function.} All $\delta$-punctured thermal states are nonclassical due to the presence of the infinite negative $\delta$-peak in the $P$ function. 

{\em Negativity of the Wigner function.} The Wigner function of $\delta$-punctured thermal states follows from Eq.~(\ref{Wfunc}) and is given by
\begin{equation}\label{Wigner1}
W(\alpha) = \frac{2\mathcal{N}}{\pi} \Bigg( \frac{e^{- \frac{2 |\alpha|^2}{1+2\bar{n}}}}{1+2\bar{n}} - w_1\, e^{-2|\alpha - \alpha_1|^2}  \Bigg).
\end{equation}
It takes negative values for large enough puncture weight 
\begin{equation}\label{NegWigner1}
w_1 > \frac{e^{- |\alpha_1|^2/{\bar{n}}}}{1+2\bar{n}}.
\end{equation}
Such weights are still acceptable as long as they do not exceed the (larger) bound (\ref{eq:b1}).

\begin{figure}[h!]
  \begin{center}
    \includegraphics[width=0.45\textwidth]{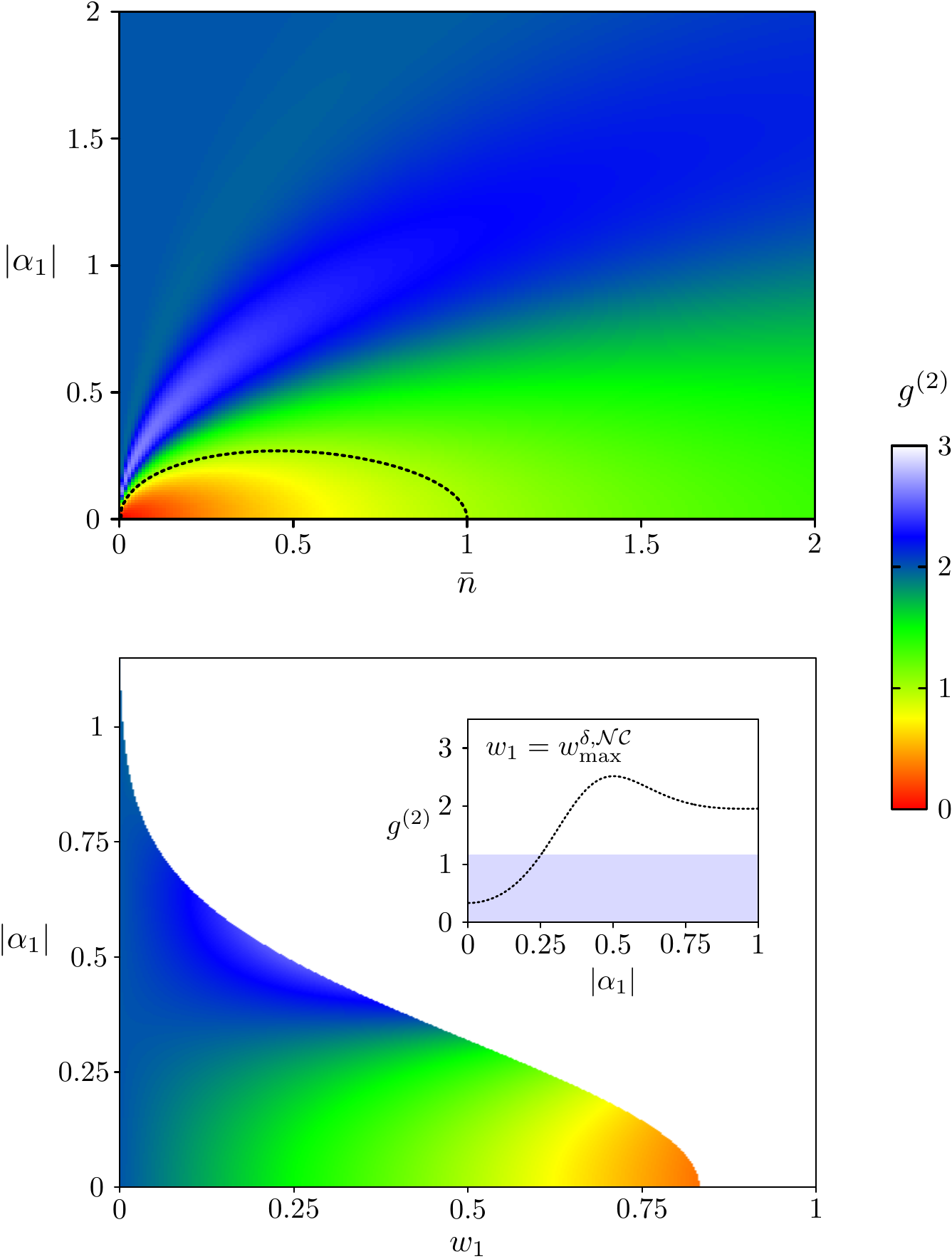}
    \caption{{Density plot of the second order correlation function
        $g^{(2)}$ [Eq.~(\ref{g2delta})] for single $\delta$-punctured
        thermal states as a function of $|\alpha_1|$ and $\bar{n}$ for
        the maximal puncture weight $\pw_1=\updelta$ given by
        Eq.~(\ref{eq:b1}) (top) and as a function of $|\alpha_1|$ and
        $\pw_1$ for $\bar{n} = 0.2$ (bottom). Only the range where
        $\rho\ge 0$ is shown. The black dashed curve in the top figure,
        along which $g^{(2)}=1$, delimits the sets of states
        displaying antibunching or not. The inset of the bottom figure
        represents $g^{(2)}$ as a function of $\alpha_1$ for the
        maximum weight $\pw_1=\updelta$. 
        }} 
    \label{fig:plotg2a}
  \end{center}
\end{figure}

{\em Antibunching.} For $\delta$-punctured thermal states, the second-order correlation function~(\ref{g2def}) is given by
\begin{equation}\label{g2delta}
g^{(2)} = \Big(1- \pw_1\Big)\; \frac{2 \bar{n}^2 - \pw_1 |\alpha_1|^4}{\left(\bar{n} - \pw_1 |\alpha_1|^2\right)^2}.
\end{equation}
For $\pw_1 \to 0$, we recover the known value $g^{(2)} = 2$ of the
thermal state~\cite{Scu01}. In Fig.~\ref{fig:plotg2a}, we show a
density plot of $g^{(2)}$ as a function of $\bar{n}$ and $|\alpha_1|$
for the maximal allowed value of the puncture weight $\pw_1$, i.e.\
the bound (\ref{eq:b1}). A whole region of parameter space corresponds
to punctured states giving rise to antibunching ($g^{(2)} < 1$). Also,
$g^{(2)}$ increases beyond 2 for $\bar n\simeq 0.1$,
$|\alpha_1|\simeq 0.5$. 

\section{Delta-punctured squeezed thermal states}

To generalize the previous results, we now replace the thermal state by a squeezed thermal state~\cite{Kim89, Bar02}
\begin{equation}\label{rhosqueezedth}
\rho_\mathrm{cl} = \frac{1}{\pi \bar{n}} \int e^{-|\alpha|^2/\bar{n}} S(r) \ket{\alpha} \bra{\alpha} S^\dagger(r)\,d^2\alpha,
\end{equation}
where $S(r)=\exp\left(-r[a^{\dagger 2} - a^{ 2}]/2 \right)$ is the squeezing operator and $r$ is the squeezing parameter, taken real for simplicity. We remind that the $P$ function of a state $\rho$ is given by the Fourier transform of its normal-ordered characteristic function $
\chi(\beta) = \mathrm{tr}[ \rho \, e^{i (\beta a^\dagger + \beta^* a)} ]$. For the squeezed thermal state $\rho_\mathrm{cl} $ defined in Eq.~(\ref{rhosqueezedth}), the normal-ordered characteristic function reads~\cite{Kim89}
\begin{equation}
\chi(\beta) = e^{-\bar{n}_I \beta_R^2 -\bar{n}_R \beta_I^2},
\end{equation}
where $\beta_R$ and $\beta_I$ are the real and imaginary parts of $\beta \in \mathbb{C}$, and 
\begin{equation}
\begin{aligned}
&\bar{n}_R = e^{-2r}\,\bar{n} - e^{-r} \,\sinh r ,\\
&\bar{n}_I = e^{2r}\, \bar{n} + e^{r}\, \sinh r .
\end{aligned}
\end{equation}
The Fourier transform of the characteristic function -- and thus the $P$ function -- exists only if both the coefficients $\bar{n}_I$ and $\bar{n}_R$ in front of $\beta_R^2$ and $\beta_I^2$ are positive, which imposes the condition
\begin{equation}\label{condChi}
e^{-2 |r|} \,\bar{n}- e^{-|r|} \sinh |r| > 0 
\end{equation} 
thereby limiting the range of allowed squeezing parameters $r\in\mathbb{R}$ for a given $\bar{n}$. The  characteristic function of a squeezed thermal state being an asymmetric Gaussian distribution, the same goes for its $P$ function. Therefore, a \emph{$\delta$-punctured squeezed thermal state} is a state determined through~(\ref{eq:rho_all}) and (\ref{eq:rho_punctured}) with
\begin{equation}\label{eq:palphaa}
\begin{aligned}
& \Pclas = \frac{e^{-\left( \frac{\alpha_R^2}{\bar{n}_R} + \frac{\alpha_I^2}{\bar{n}_I}\right)}}{\pi \sqrt{\bar{n}_R \bar{n}_I}} , \, \\[2pt]
& \pi(\alpha)=\delta(\alpha),
\end{aligned}
\end{equation}
where $\alpha_R$ and $\alpha_I$ are the real and imaginary parts of $\alpha
\in \mathbb{C}$.

\subsection{Conditions for (semi)positive definiteness}

{\em Necessary condition.}
Since $\Pclas$ is rotationally non-symmetric, it is reasonable to assume that the necessary condition~(\ref{NC}) will provide us with a tight bound for $w_1$ only if we consider states $\ket{\phi}\in \mathcal{K}$ with non-symmetric $P-$functions. By taking $\mathcal{K}$ to be the set of squeezed coherent states~\cite{Kim89,Bar02}, we obtain the following upper bound for the puncture weight $w_1$:
\begin{equation}
{
\pw_1 \leqslant \frac{2 \sqrt{\bar{n}_R \bar{n}_I} \, \, e^{-\left(\frac{\alpha_{1R}^2}{\bar{n}_R} + \frac{\alpha_{1I}^2}{\bar{n}_I}\right)}}{\bar{n}_R + \bar{n}_I + 2 \bar{n}_R \bar{n}_I}\equiv \updeltasqueezed .
}
\label{eq:b1as}
\end{equation}
When $\bar{n}_R=\bar{n}_I=\bar{n}$, this condition reduces to the condition (\ref{eq:b1}), as expected.

\begin{proof}
Let us denote by $\ket{\gamma, r'}$ a squeezed coherent state defined as~\cite{Kim89,Bar02}
\begin{equation}\label{sqstate}
\ket{\gamma, r'} = D(\gamma) S(r') \ket{0},
\end{equation}
where $D(\gamma)= \exp\left(\gamma a^\dagger - \gamma^* a\right) $ is the displacement operator and $S(r')$ the squeezing operator of parameter $r'$, where the prime symbol distinguishes $r'$ from $r$, this latter defining $\bar{n}_R$ and $\bar{n}_I$.
The necessary condition~(\ref{NC}) with $\ket{\phi} = \ket{\gamma, r'}$ reads
\begin{equation}\label{NCsquproof}
\pw_1 \leq  \frac{\int \Pclas |\langle \gamma, r' |\alpha\rangle|^2  d^2 \alpha }{|\langle \gamma, r' |\alpha_1\rangle|^2},
\end{equation}
where the modulus squared of the scalar product between a squeezed coherent state $\ket{\gamma, r'}$ and a coherent state
$\ket{\alpha}$ is given by~\cite{Scu01}
\begin{multline}
\left| \langle \gamma, r' | \alpha \rangle \right|^2 =
\mathrm{sech}(r') \, e^{-(|\alpha|^2 + |\gamma|^2) + (\alpha^* \gamma + \gamma^* \alpha) \mathrm{sech}(r')} \\ \times e^{- \mathrm{th}(r')\, \mathrm{Re}
    \big[\alpha^2 - \gamma^2\big]}.
\end{multline}
The minimum of the rhs term in Eq.~(\ref{NCsquproof}) is obtained, after integration over $\alpha$, for
\begin{equation}\label{gammarsol}
\begin{aligned}
&\gamma = \frac{\bar{n}_R + \bar{n}_I + 2 \bar{n}_R \bar{n}_I}{\sqrt{\left(\bar{n}_R - \bar{n}_I + 2 \bar{n}_R \bar{n}_I\right)\left(\bar{n}_I - \bar{n}_R + 2 \bar{n}_R \bar{n}_I\right)}} \, \alpha_{1},\\
&r' = \mathrm{arctanh}\left(\frac{\bar{n}_R-\bar{n}_I}{2\bar{n}_R \bar{n}_I}\right)
\end{aligned}
\end{equation} 
and leads to the necessary condition (\ref{eq:b1as}).
\end{proof} 

\begin{figure}[h!]
  \begin{center}
    \includegraphics[width=0.45\textwidth]{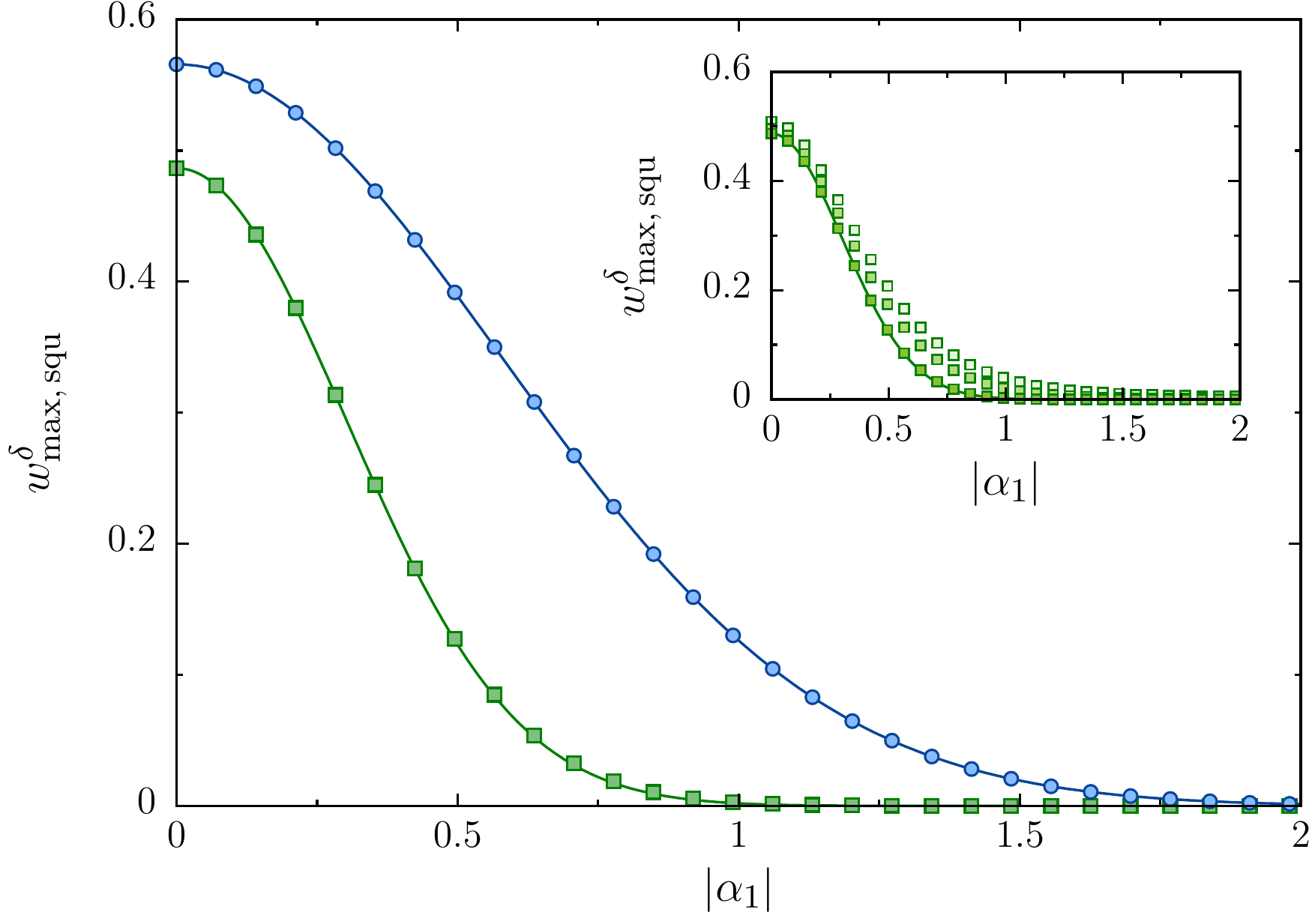}
    \caption{Plot of the upper bound $\updeltasqueezed$ (solid curves) and of $\pw_{\mathrm{max, \, squ}}^\delta$, the $\pw_1$ which cancels the smallest eigenvalue of the density matrix, as a function of $|\alpha_1|$ for $\bar{n}_R=1$, $\bar{n}_I=0.1$ (squares) and $\bar{n}_I=0.5$ (dots). The eigenvalues are computed in a Fock space of dimension $n_\text{max} = 25$. Inset: convergence of $\pw_{\mathrm{max}, \, \mathrm{squ}}^\delta$ (green squares) with increasing cut-off dimension $n_\mathrm{max}$ ($n_\mathrm{max}=6,8,25$ from top to bottom) for $\bar{n}_I=0.1$.} 
    \label{fig:tightsq}
  \end{center}
\end{figure}

{\em Tightness of $\updeltasqueezed$.}
Computations relying on the diagonalization of $\rho$ given by Eq.~(\ref{eq:rho_all}) with Eqs.~(\ref{eq:rho_punctured}) and~(\ref{eq:palphaa}) suggest that the
bound $\updeltasqueezed$ given in Eq.~(\ref{eq:b1as}) is tight. As for the case of thermal states, this claim is supported by the observation that the squeezed coherent state $\ket{\gamma, r'}$ with $\gamma$ and $r'$ given by Eq.~(\ref{gammarsol}) is an eigenstate of $\rho$ with eigenvalue $0$.
Figure \ref{fig:tightsq} shows $\updeltasqueezed$ and the value of $\pw_1$ that cancels the smallest eigenvalue of $\rho$. Numerical results converge to the bound as the cut-off dimension increases.

\subsection{Conditions for nonclassicality}

{\em Negativity of the $P$ function.} As for $\delta$-punctured thermal states, the presence of a $\delta$ peak always enforces nonclassicality.

{\em Negativity of the Wigner function.} The Wigner function of a $\delta$-punctured squeezed thermal state reads
\begin{equation}
W(\alpha) = \frac{2\mathcal{N} }{\pi} \Bigg( \frac{e^{- 2 \big( \frac{\alpha_R^2}{1 + 2 n_R} + \frac{\alpha_I^2}{1+2n_I}\big)}}{1+2\bar{n}} - w_1\,e^{-2|\alpha - \alpha_1|^2}  \Bigg),
\end{equation}
and takes negative values if the puncture weight satisfies
\begin{equation}
w_1 > \frac{e^{-\left(\frac{\alpha_R^2}{\bar{n}_R}+\frac{\alpha_I^2}{\bar{n}_I}\right)}}{1+2\bar{n}},
\end{equation}
which is still acceptable as long as it does not exceed the (larger) bound (\ref{eq:b1as}).

\begin{figure}[h!]
  \begin{center}
    \includegraphics[width=0.45\textwidth]{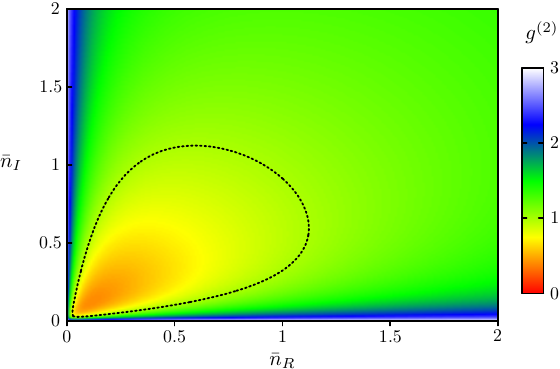}
    \caption{{Density plot of the second order correlation function $g^{(2)}$ [Eq.~(\ref{g2deltasqueezed})] for single $\delta$-punctured squeezed thermal states as a function of $\bar{n}_R$ and $\bar{n}_I$ for the maximal puncture weight $\pw_1$ given in Eq.~(\ref{eq:b1as}) for $\alpha_{1R} = \alpha_{1I} = 0.1/\sqrt{2}$. The black dashed curve delimits the region where $g^{(2)} < 1$. Note that in the absence of puncture, $g^{(2)} > 2$ for any squeezed thermal state (see text).}} 
    \label{fig:plotg2b}
  \end{center}
\end{figure}

{\em Antibunching.} For $\delta$-punctured squeezed thermal states, the second-order correlation function~(\ref{g2def}) is given by
\begin{equation}\label{g2deltasqueezed}
g^{(2)} = \Big(1-\pw_1\Big)\; \frac{  \frac{1}{4}\left(3 \bar{n}_R^2 + 2 \bar{n}_R \bar{n}_I + 3 \bar{n}_I^2 \right) - \pw_1 |\alpha_1|^4}{\left(\frac{\bar{n}_R+ \bar{n}_I}{2} - \pw_1 |\alpha_1|^2\right)^2}.
\end{equation}
For $\pw_1 \to 0$, we recover the known value for squeezed thermal states~\cite{Kim89}
\begin{equation}
\begin{aligned}
g^{(2)} &= \frac{3 \bar{n}_R^2 + 2 \bar{n}_R \bar{n}_I + 3 \bar{n}_I^2}{(\bar{n}_R + \bar{n}_I)^2} \\
&= 2 + \frac{(2\bar{n} + 1)^2 \sinh^2r \cosh^2r  }{\big(\bar{n} \cosh(2r)+ \sinh^2 r\big)^2} \geqslant 2.
\end{aligned}
\end{equation}
Figure~\ref{fig:plotg2b} shows a density plot of $g^{(2)}$ as a function of $\bar{n}_R$ and $\bar{n}_I$ for $|\alpha_1|=0.1$ with the maximal allowed value of $\pw_1$, i.e.\ the bound (\ref{eq:b1as}). Again, a whole region of parameter space (delimited by the black dashed curve) corresponds to states displaying antibunching ($g^{(2)} < 1$), to be contrasted with squeezed thermal states (i.e.\ without puncture) for which $g^{(2)} \geqslant 2$. 

\section{Gaussian-punctured thermal states}

We now generalize the results of Sec.~\ref{sec:deltap} by replacing $\delta$-punctures by single narrow Gaussian punctures. We thus define \emph{Gaussian-punctured thermal states} as states determined through~(\ref{eq:rho_all}) and (\ref{eq:rho_punctured}) with
\begin{equation}\label{def:general_state}
\begin{aligned}
& \Pclas =  \frac{e^{-|\alpha|^2/\bar{n}}}{\pi
  \bar{n}}, \, \\
& \pi(\alpha)= \frac{e^{-|\alpha|^2/b}}{\pi
  b},
\end{aligned}
\end{equation}
where $b > 0$ characterizes the width of the puncture. Note that $\delta$-punctured thermal states correspond to the limit $b \to 0$.

\subsection{Vacuum-centered Gaussian punctures}\label{sec:VCGP}
We first consider the simple case of a thermal state from which we
subtract a single Gaussian centered at $\alpha_1=0$. A thermal state with the vacuum component exactly removed (case $b\to 0, \bar{n} = 1, w_1=1/2$) has already been considered in the literature as a simple example of a nonclassical state \cite{Dio00}.  Here we keep a variable width and amplitude for the subtracted Gaussian.

\subsubsection{Condition for (semi)positive definiteness}

{\em Necessary and sufficient condition.} Since both
the original and the subtracted state are diagonal in the Fock state
basis, $\mathcal{NSC}$ can be tackled analytically. The $P$ function
of the vacuum-centred Gaussian-punctured thermal state 
reads 
\begin{equation}\label{PvcG}
P(\alpha) = \mathcal{N} \left(
  \frac{e^{-\frac{|\alpha|^2}{\bar{n}}}}{\pi \bar{n}}-  \pw_1
  \frac{e^{-\frac{|\alpha|^2}{b}}}{\pi b}  \right). 
\end{equation}
It takes negative values at $\alpha=0$ as soon as
$\pw_1>b/\bar{n}$. As we will show, only $b<\bar{n}$ can lead to a
positive semidefinite density operator. 
Expressing the coherent states in the Fock state basis using Eq.~(\ref{alphaasFock}) and performing the
Gaussian integrals, we find
\begin{equation*}
\rho = \mathcal{N} \sum_{n=0}^\infty \big [ p_{\mathrm{cl}}(n) - \pw_1\,p_\pi(n) \big ] \ket{n}\bra{n},
\end{equation*}
with
\begin{equation}
\begin{aligned}
p_{\mathrm{cl}}(n) = \frac{\bar{n}^n}{(\bar{n}+1)^{n+1}},\quad
p_\pi(n) = \frac{b^n}{(b+1)^{n+1}}.
\end{aligned}
\end{equation}
Since $\rho$ is diagonal, the positivity condition reads $p_{\mathrm{cl}}(n) - \pw_1\,p_\pi(n) \geq 0$ for all $n\in \mathbb{N}_0$, which is equivalent to
\begin{equation}
\pw_1 \leq \frac{b+1}{\bar{n}+1} \inf_{n\in \mathbb{N}_0} \left\lbrace\left(\frac{\bar{n}b+\bar{n}}{\bar{n}b+b}\right)^n\right\rbrace  \equiv \nscgaussvac.
\end{equation}

We can now distinguish two cases: i) If $b> \bar{n}$, then the fraction in the argument of the infimum is smaller than 1 and the infimum is obtained in the limit $n \rightarrow \infty$ and evaluates to $0$, hence $w_1=0$. This shows that we cannot subtract a broader Gaussian than the original one. ii) In the other case, where $b\leqslant\bar{n}$, the minimum is obtained
  for $n=0$ and evaluates to 1. We can thus subtract a Gaussian that
  is tighter than the original one and still obtain a positive semidefinite
  density operator. This works as long as 
\begin{align*}
\pw_1 \leq \frac{b+1}{\bar{n}+1} \equiv \nscgaussvac.
\end{align*}

\subsubsection{Conditions for nonclassicality}

{\em Negativity of the $P$ function.} In order to obtain a $P$ function which exhibits negative values for some $\alpha$, we see from Eq.~(\ref{PvcG}) that this requires 
\begin{equation}\label{SCnonclP}
w_1 > \frac{b}{\bar{n}},
\end{equation}
showing that in the limit $b \to 0$, we recover the fact that all $\delta$-punctured states are nonclassical. 

{\em Negativity of the Wigner function.} The Wigner function of a vacuum-centered Gaussian-punctured thermal state reads
\begin{equation}\label{Wigner3}
W(\alpha) = \mathcal{N} \frac{2}{\pi} \left( \frac{e^{- \frac{2 |\alpha|^2}{1+2\bar{n}}}}{1+2\bar{n}} - w_1 \frac{e^{- \frac{2 |\alpha|^2}{1+2 b}}}{1+2 b} \right),
\end{equation}
and takes negative values if $w_1$ satisfies
\begin{equation}\label{NegWignerG0}
w_1 > \frac{1+2 b}{1+2 \bar{n}},
\end{equation}
which is a stronger condition than Eq.~(\ref{SCnonclP}). Note that for $b \to 0$, Eq.~(\ref{NegWignerG0}) tends to Eq.~(\ref{NegWigner1}), the bound found for $\delta$-punctured thermal states.

{\em Antibunching.} From the expression of the second-order correlation function~(\ref{g2def}), we have
\begin{equation}\label{g2gaussvac}
\begin{aligned}
g^{(2)} = (1-\pw_1) \frac{2\bar{n}^2-\pw_1 2b^2 }{\left(\bar{n}-\pw_1 b \right)^2}.
\end{aligned}
\end{equation}
For $w_1 = \nscgaussvac$, the condition for antibunching reads 
\begin{equation}
g^{(2)}<1\quad\Leftrightarrow\quad \bar{n}^2 + b^2 < 1,
\end{equation}

To summarize, for $\alpha_1=0$, a $P$ function that corresponds to a positive semidefinite density
operator and attains negative values in some region is obtained if and only if 
\begin{align}
{
b< \bar{n} \qquad \text{and} \qquad \frac{b}{\bar{n}} < \pw_1 \leq \frac{b+1}{\bar{n}+1}.
}
\label{eq:bounds_vacuum}
\end{align}
Hence, the puncture weight $w_1$ must be smaller than $(b+1)/(\bar{n}+1)$ to ensure the physicality of the state, but greater than $b/\bar{n}$ to yield a nonclassical state. Moreover, in the limit $b \rightarrow 0$, the Gaussian puncture becomes a
$\delta$-puncture, and we recover the value given in \eqref{eq:b1}. This provides an analytical proof of the tightness of the bound $\updelta$ in the case $\alpha_1 =0$.

\subsection{Arbitrarily centred Gaussian punctures}
\label{acGpsec}

We now consider the more general case of an
arbitrarily centred Gaussian puncture corresponding to a $P$ function of the form
\begin{equation}\label{PG}
P(\alpha) = \mathcal{N} \left( \frac{e^{-\frac{|\alpha|^2}{\bar{n}}}}{\pi \bar{n}}-  \pw_1\frac{e^{-\frac{|\alpha - \alpha_1|^2}{b}}}{\pi b}  \right).
\end{equation}
We first obtain an upper bound on the
amplitude of the puncture, and then support the tightness of this
bound with numerical results. Finally, we find the regimes of parameters yielding nonclassicality and antibunching.

\subsubsection{Condition for (semi)positive definiteness}
\label{sec:tight}

{\em Necessary condition.} Using Eq.~(\ref{NC}) with $\mathcal{K}$ the set of coherent states, we find that we
can only subtract a Gaussian tighter than the original state
($b<\bar{n}$) and obtain as a necessary condition for positivity
\begin{equation}
{
\pw_1 \leq \frac{b+1}{\bar{n}+1} e^{-\frac{|\alpha_1|^2}{\bar{n}-b}} \equiv \upncgauss.
}
\label{eq:cup_coherent}
\end{equation}
\begin{proof}
Using $|\langle\gamma|\alpha\rangle|^2 = e^{-|\alpha-\gamma|^2}$ and Eq.~\eqref{def:general_state}, we find for the expectation value of $\rho$ in the coherent state $\ket{\gamma}$ 
\begin{equation}
\begin{aligned}
\rho_{\gamma\gamma} = \int \Big(\frac{1}{\pi \bar{n}}e^{-\frac{|\alpha|^2}{\bar{n}}-|\alpha-\gamma|^2} - \frac{\pw_1}{\pi b} e^{-\frac{|\alpha-\alpha_1|^2}{b}-|\alpha-\gamma|^2}\Big)d^2\alpha.
\end{aligned}
\end{equation}
Evaluation of the Gaussian integrals leads to
\begin{equation}
\rho_{\gamma\gamma} \geq 0\quad\Leftrightarrow\quad\pw_1 \leq \frac{b+1}{\bar{n}+1}e^{-\frac{|\gamma|^2}{\bar{n}+1}+\frac{|\gamma-\alpha_1|^2}{b+1}}.\label{eq:c}
\end{equation}
That needs to be true for all $\gamma\in\mathbb{C}$ in order to keep the
possibility of a positive semidefinite density operator. Minimizing this expression
over $\gamma$ gives us the upper bound $\upncgauss$. Since the exponential function is strictly monotonically increasing and the prefactors are positive we
can focus on minimizing the exponent: 
\begin{align*}
f(\gamma) \equiv -\frac{|\gamma|^2}{\bar{n}+1}+\frac{|\gamma-\alpha_1|^2}{b+1}.
\end{align*}
By applying a rotation in phase-space, the puncture's center can always be brought along the real axis, so that we can set $\alpha_1 = x_1 \in
\mathbb{R}$. We then split $\gamma$ into its real and imaginary part
$\gamma = \gamma_R + i\gamma_I$, leading to 
\begin{align*}
f(\gamma)  = h(\gamma_I) + g(\gamma_R) + \frac{x_1^2}{b+1}
\end{align*}
with
\begin{align*}
h(\gamma_I)  &= \gamma_I^2 \left(\frac{1}{b+1}-\frac{1}{\bar{n}+1}\right),  \\
g(\gamma_R) &= \gamma_R^2\left(\frac{1}{b+1}-\frac{1}{\bar{n}+1}\right) - \frac{2\gamma_Rx_1}{b+1}.
\end{align*}

Here we have to distinguish two cases (the case $b=\bar{n}$ is trivial): i) If $b>\bar{n}$, $h(\gamma_I)$ is an inverted
parabola and can attain arbitrary negative values and the same holds for
$f$, thus pushing the exponential in (\ref{eq:c}) arbitrary close to
$0$, implying that the above condition can only be fulfilled for all 
coherent states if $\pw_1=0$. This means that we cannot subtract any
Gaussian wider than the original thermal one and still obtain a valid
state. This generalizes what we have already seen in the
case of the vacuum centred Gaussian. ii) For $b<\bar{n}$, the dependence of $f$ on the parameters $\gamma_R,\gamma_I$ splits into the two
functions $h$ and $g$ that depend on different independent arguments
$\gamma_R$ and $\gamma_I$. Thus we can minimize both $h$ and $g$
separately. Obviously $h$ is minimized for $\gamma_I=0$. We find that $g(\gamma_R)$ is minimized for $\gamma_{R,\text{min}} = (\bar{n}+1) x_1/(\bar{n}-b)$ and we get the minimal values
\begin{align*}
& g(\gamma_{R,\text{min}}) = -x_1^2 \frac{\bar{n}+1}{b+1}\frac{1}{\bar{n}- b},\\
& f(\gamma_{R,\text{min}}) = -\frac{x_1^2}{\bar{n}-b}.
\end{align*}
Finally, for the general case $\alpha_1 \in \mathbb{C}$, plugging $f(\gamma_{R,\text{min}})$ into Eq.~(\ref{eq:c}) yields the bound (\ref{eq:cup_coherent}).
\end{proof}

{\em Tightness of $\upncgauss$.} \label{sec:numerical} 
As for the previous bounds obtained from the necessary condition~(\ref{NC}), diagonalizing the density operator of the Gaussian punctured thermal state  [i.e.\ Eq.~(\ref{eq:rho_all}) together with Eqs.~(\ref{eq:rho_punctured}) and (\ref{def:general_state})] and using the $\mathcal{NSC}$ suggest that the bound $\upncgauss$ [Eq.~(\ref{eq:cup_coherent})] is tight. This result is shown in Fig.~\ref{b1nbar1,5}, where we plotted the analytic bound (lines) and the numerical bound (squares and dots) for different cut-off dimensions $n_\text{max}$ as a function of the center of the puncture $|\alpha_1|$. 
Also,  we found that the coherent state $\ket{\gamma}$ with $\gamma = (\bar{n}+1)/(\bar{n}-b)\, \alpha_1$ is an eigenstate of $\rho$ with eigenvalue $0$ when $w_1$ is given by the bound~(\ref{eq:cup_coherent}).
\begin{figure}
\includegraphics[scale=0.475]{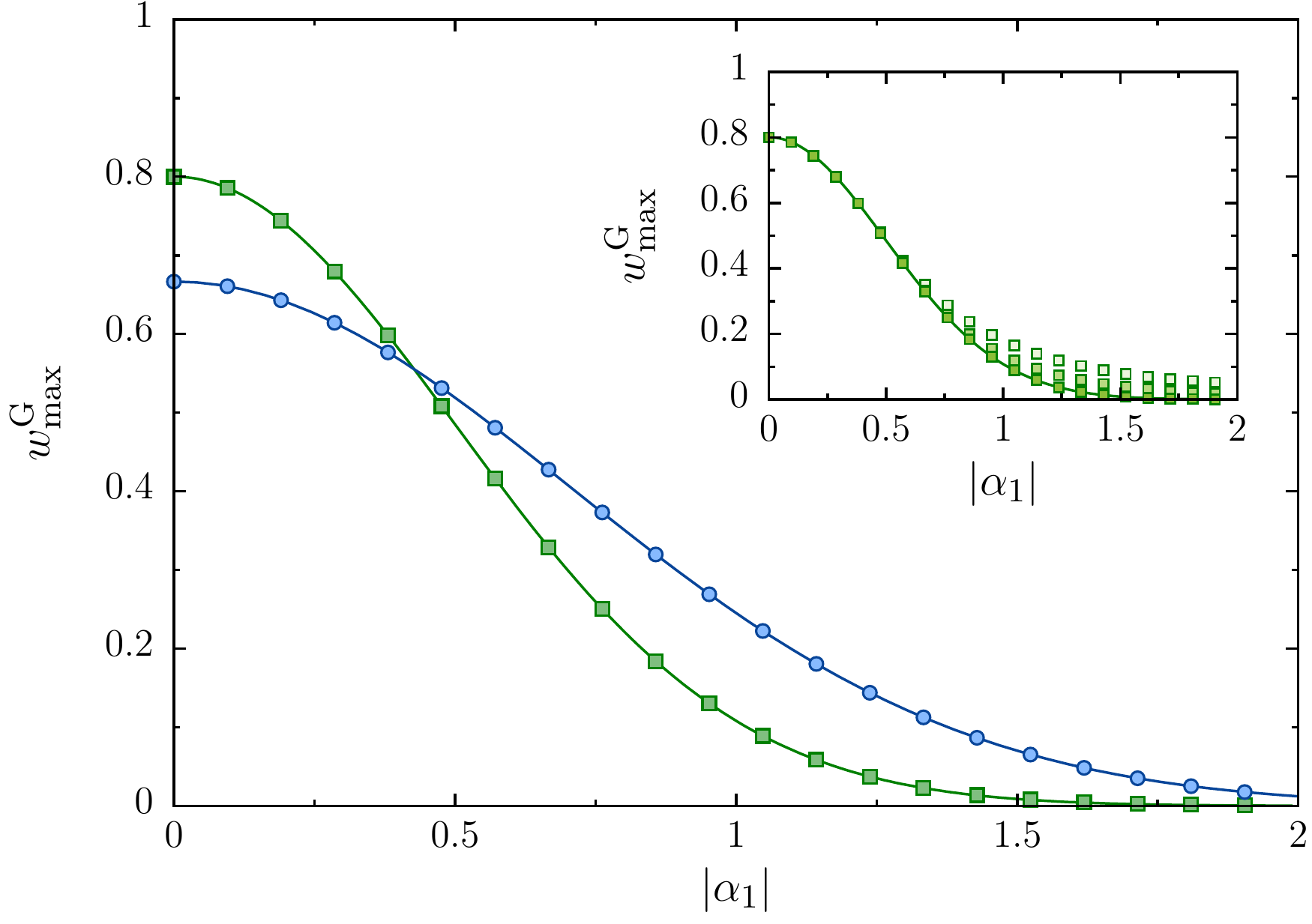}
\caption{Plot of the upper bound $\upncgauss$ (solid curve) and of $\pw_{\mathrm{max}}^\text{G}$, the $\pw_1$ which cancels the smallest eigenvalue of the density matrix, as a function of $|\alpha_1|$ for $b=1$ and $\bar{n}=1.5$ (squares) and $\bar{n}=2$ (dots). The eigenvalues are computed in a Fock space of dimension $n_\text{max} = 40$. Inset: convergence of $\pw_{\mathrm{max}}^\text{G}$ (green squares) as a function of the cut-off dimension $n_\mathrm{max}$ ($n_\mathrm{max}=8,12,40$ from top to bottom).} 

\label{b1nbar1,5}
\end{figure}

\subsubsection{Conditions for nonclassicality}
\label{Gacp}

{\em Negativity of the $P$ function.} From Eq.~(\ref{PG}), we see that partly negative $P$ functions can be obtained when 
\begin{equation}\label{SCnonclParbi}
w_1 > \frac{b}{\bar{n}}\, e^{-\frac{|\alpha_1|^2}{\bar{n}-b}},
\end{equation}
which generalizes Eq.~(\ref{SCnonclP}). Hence, the lower bound of the
weight decreases with the distance between the center of the thermal state and the position of the puncture. Further this is smaller than $\upncgauss$ given in \eqref{eq:cup_coherent} since $\bar{n}>b$, implying that for all allowed values of $\alpha_1$, $\bar{n}$ and $b$, there is a puncture weight such that the $P$ function is negative somewhere in the complex plane, and the state is nonclassical.

{\em Negativity of the Wigner function.} The Wigner function of an arbitrarily-centred Gaussian-punctured thermal state reads
\begin{equation}\label{Wigner4}
W(\alpha) = \frac{2\mathcal{N}}{\pi} \left( \frac{e^{- \frac{2 |\alpha|^2}{1+2\bar{n}}}}{1+2\bar{n}} - w_1 \frac{e^{- \frac{2 |\alpha - \alpha_1|^2}{1+2 b}}}{1+2 b} \right),
\end{equation}
and is negative if $w_1$ satisfies
\begin{equation}\label{NegWigner4}
w_1 > \frac{1+2b}{1+2\bar{n}}\, e^{- \frac{|\alpha_1|^2}{\bar{n}-b}},
\end{equation}
a stronger condition than Eq.~(\ref{SCnonclParbi}).

\begin{figure}[h!]
  \begin{center}
    \includegraphics[width=0.45\textwidth]{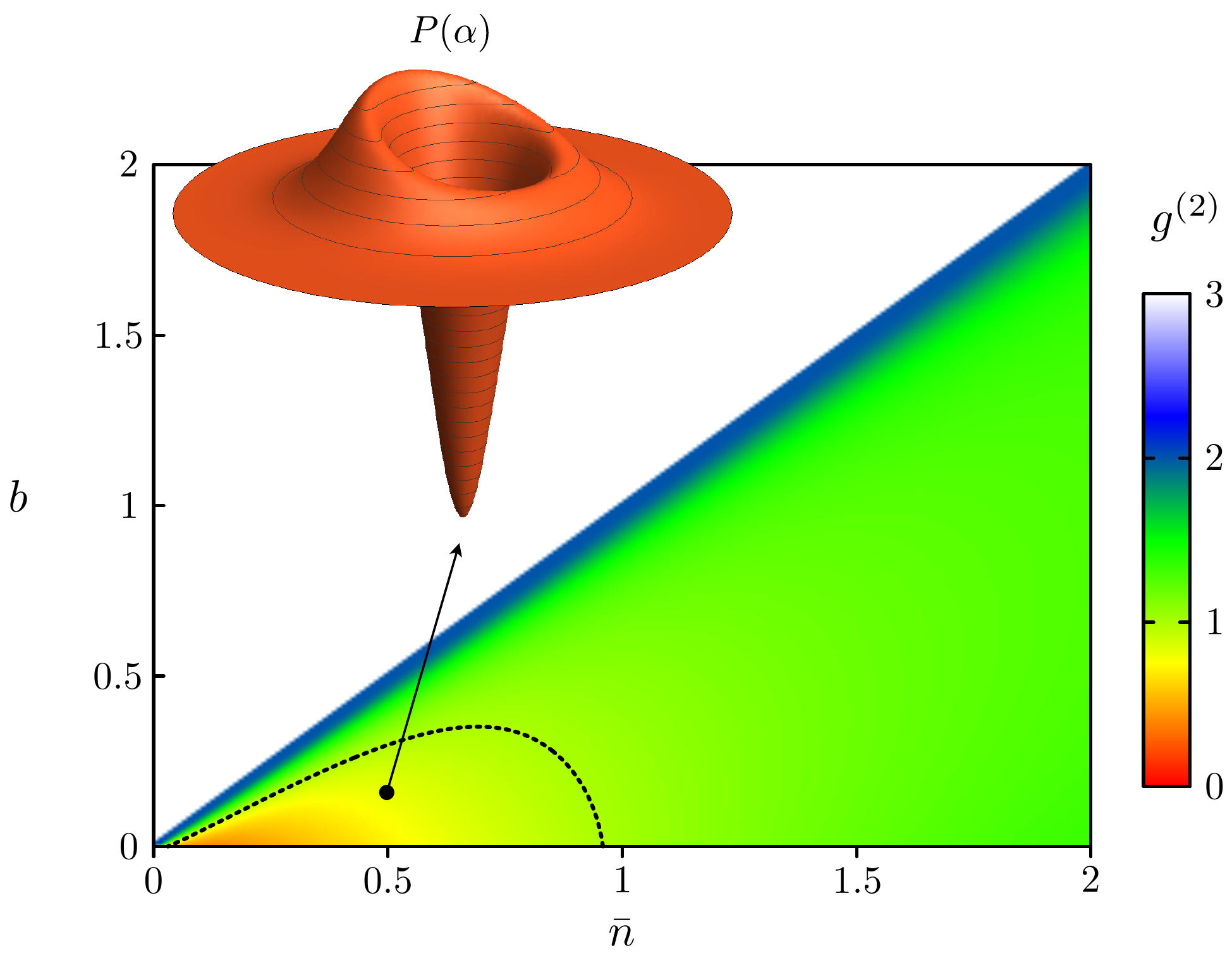}
    \caption{{Density plot of the second order correlation function $g^{(2)}$ as a function of $\bar{n}$ and $b<\bar{n}$ for single Gaussian-punctured states with $\alpha_1=0.1$
for the maximal puncture weight $\pw_1$ given in Eq.~(\ref{eq:cup_coherent}). The black dashed curve delimits the region where $g^{(2)} < 1$. In the top left corner is shown the $P$ function for the parameters indicated by a dot in the density plot.
        }} 
    \label{fig:plotg2_gaus}
  \end{center}
\end{figure}

{\em Antibunching.} The correlation function~(\ref{g2def}) for the state \eqref{PG} is given by
\begin{equation}\label{g2gauss}
\begin{aligned}
g^{(2)} = (1-\pw_1) \frac{2\bar{n}^2-\pw_1\left(|\alpha_1|^4 + 4 |\alpha_1|^2 b+2b^2\right)}{\left[\bar{n}-\pw_1(|\alpha_1|^2+b)\right]^2}.
\end{aligned}
\end{equation}
Note that in the limit $b\rightarrow 0$, Eq.~(\ref{g2gauss}) tends to Eq.~\eqref{g2delta} found in Sec.~\ref{sec:deltap} for a $\delta$-punctured thermal state. We have that for $\bar{n}< 1$, the minimum of $g^{(2)}$ is obtained for $b \rightarrow 0$. For $\bar{n}\rightarrow 0 $ this minimum tends to $0$. We also have that for $\bar{n} > 1$, the minimum of $g^{(2)}$ is obtained for a finite value of $b$. Figure~\ref{fig:plotg2_gaus} shows Eq.~(\ref{g2gauss}) as a function of $\bar{n}$ and $b<\bar{n}$ for $\alpha_1 =0.1$ and for the maximal puncture weight $\pw_1 = \upncgauss$ given in Eq.~\eqref{eq:cup_coherent}.

\section{Possible experimental realizations of punctured states}

\subsection{Vacuum-removed states}
A $\delta$-punctured thermal state with puncture at ${\alpha_1=0}$ and
maximum weight of the puncture,
$b_\mathcal{NSC}^\text{gaus,vac}$, i.e.~a vacuum-removed thermal
state, 
\begin{equation}
  \label{eq:desired}
\rho = \mathcal{N} \sum_{n=1}^\infty \frac{\bar{n}^n}{(1+\bar{n})^{n+1}}\ket{n}\bra{n} = \frac{\bar{n}+1}{\bar{n}}\left[\rho_T -\frac{1}{1+\bar{n}}\ket{0}\bra{0}\right],  
\end{equation}
is probably the example of 
a punctured state that is easiest to realize experimentally. 
Since a thermal
state has no coherences between Fock states, a projective
measurement in the Fock-basis does not destroy the quantum properties
of the state, but only alters its  statistical character. Discarding
the ground-state when it is found leads to an ensemble of states that
realizes the vacuum-removed thermal state.  This generalizes to any
initial state that is diagonal in the Fock-basis.  When single-photon sources are
available, one can of course synthesize a given vacuum-removed state
diagonal in the Fock basis from the
beginning by mixing Fock-states with the corresponding statistical
weight without the need for any measurement. This generalizes to
states where the vacuum is not removed completely by adjusting the
statistical weight of the vacuum-state to the desired value.\\

\begin{figure*}
\centering
\includegraphics{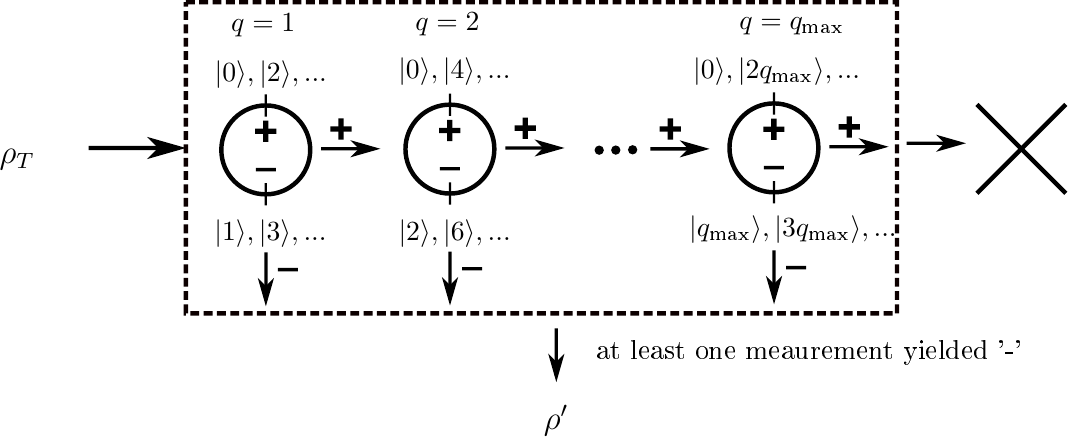}
\caption{Sketch of realization of vacuum-removed thermal state. A
  thermal state $\rho_T$ is subjected to multiple
  QND-measurements that distinguish the vacuum state of other photon number states. If at
  least one of the measurements yields $'-'$, the vacuum has definitely not been
  realized and the measured state is added to the ensemble. If all
  measurements yielded $'+'$ the state is
  discarded. Since we only perform a
  finite number of measurements, the final state
  $\rho'$ given in equation (\ref{eq:realized}) can slightly deviate
  from the   desired state $\rho$ given in equation (\ref{eq:desired}). The
  deviation can be quantified calculating the fidelity
  (\ref{eq:fidelity}). } 
\label{fig:exp}
\end{figure*}

The measurement-based 
procedure faces the problem that
traditional photon-counting methods destroy the photons during
the measurement. 
Here we propose a procedure suitable for
cavity-QED or circuit-QED that realizes a vacuum-removed state
diagonal in the Fock basis, based on the nondestructive detection of photons \cite{Rei13, Gab92, Gue07, Xia08, Hel09} in a mode of the cavity (See Fig.~\ref{fig:exp}). Two-level atoms in the cavity in a superposition 
$\ket{+}$ of their two states 
$\ket{g},\ket{e}$, where $\ket{\pm}=(\ket{g}\pm\ket{e})/\sqrt{2}$,
experience a phase shift $\varphi = \frac{\pi (n\mod 
  2q)}{q}$ between $\ket{g}$ and $\ket{e}$ that depends on the number
of photons in the cavity, with 
an  integer $q$ that can be controlled through the interaction time
and strength. 
A projective measurement of an atom in the basis $\{\ket{\pm}\}$
(corresponding to measurements results labelled $\pm$)
updates our knowledge of the photon number.  
Since $\varphi$ is defined modulo $2\pi$,  
$2q$ different Fock states can be distinguished.\\ 

However, we need no information which Fock state is actually 
realized, but only need to systematically exclude the vacuum state. This
can be achieved with an iterative procedure: In the first step one sets
$q=1=2^0$. If '$-$' is 
measured,  the state is tagged as part of the ensemble. If '$+$' is
measured, we  
still need to distinguish the states $\ket{n}$ with $n= 0, 2, 4, 6,
8,...$ in order to not truncate more states than the vacuum.  
So we set $q=2= 2^1$ yielding a phase shift of 
 $\varphi = \frac{\pi (n \mod 4)}{2}$ 
that allows us to distinguish states with $n = 2+4k, \; k\in
\mathbb{N}_0$ photons (measurement result ``-'') from states with $n = 4k, \; k\in
\mathbb{N}_0$, including the vacuum (measurement result ``+''). 
This can be continued with $q = 2^{l-1},\; l = 3,4,...,
l_\text{max}$, as long as needed to exclude the ambiguity of states
with even photon numbers as
high as wished.\\ 
In general, setting $q=2^{l-1}$, with
$l\in\mathbb{N}$, we can definitely distinguish the vacuum from all
states with $n = q +2kq, k \in \mathbb{N}_0$.  
If after a finite number of repetitions $l_\text{max}$ all
measurements yielded $'+'$ we
need to completely 
discard the remaining state.  This obviously causes some errors, since
we also exclude possible realizations other than the vacuum from being
further used. The smallest photon number that was not distinguished
from the vacuum, and was therefore also subtracted in the remaining
state, is then $n_0 = 2 q_\text{max}=2^{l_\text{max}}$, where
$q_\text{max}$ corresponds to the smallest interaction time that has
been realized, i.e. the last measurement.  
The whole procedure post-selects states $\ket{n} \neq
\ket{2^{l_\text{max}}k} \forall  k\in \mathbb{N}_0$, whereas states
$\ket{n} = \ket{2^{l_\text{max}}k}, \;  k\in \mathbb{N}_0$ are
truncated. The procedure only
works because 
the thermal state is diagonal in the Fock-basis; all coherences
between different Fock states are lost.  In such a case, however, the
procedure is very effective: If we want to make sure that
states with photon number less than $n_0$ 
were not truncated, we only need to perform $l_\text{max} =
\lceil\log_2 (n_0)\rceil$ different measurements.\\ 

For a thermal state, occupation for high photon number
states decays very rapidly for small temperatures, and the overall error can be kept
small. To quantify this, we can calculate the fidelity of our desired state \eqref{eq:desired} with respect to the  state that was actually constructed
\begin{align}
\rho' &=\mathcal{N}'\left[\rho_T - \sum_{n=0}^\infty \frac{\bar{n}^{2^{l_\text{max}}n}}{(1+\bar{n})^{2^{l_\text{max}}n+1}}\ket{2^{l_\text{max}}n}\bra{2^{l_\text{max}}n}\right]\label{eq:realized}.
\end{align}
Since both density
operators are diagonal in Fock basis the fidelity is simply evaluated
as \cite[p. 409]{Nie10}: 
\begin{align}
\begin{aligned}
F(\rho,\rho') &= \sum_{n=0}^\infty \sqrt{ \rho_{nn} \rho_{nn}'} =  1 - \frac{1}{\bar{n}}\frac{1}{\left(\frac{\bar{n}+1}{\bar{n}}\right)^{2^{l_\text{max}}}-1}. \label{eq:fidelity}
\end{aligned}
\end{align}
In the limit of large $l_\text{max}$ this reduces to
\begin{align}
F = 1- \frac{1}{\bar{n}}\left(\frac{\bar{n}}{\bar{n}+1}\right)^{2^{l_\text{max}}} = 1- \frac{e^{-\log(\frac{\bar{n}+1}{\bar{n}})2^{l_\text{max}}}}{\bar{n}},
\end{align}
i.e.~the fidelity approaches $1$ exponentially with
$2^{l_\text{max}}$.

This shows the theoretical viability of the procedure.  Experimental
imperfections, as for example the non perfect linearity of the phase
shift in the photon numbers~\cite{Gue07}, may worsen the result and
will have to be evaluated for a given experimental setup.

\subsection{``Vacuum or not''-measurements} 
The vacuum-state can be removed from a broader class of states, namely
states that do not contain coherences between the vacuum
state and other Fock states.
Indeed, if $\rho_{nn}=0$ in the Fock
basis, then $\rho_{0n}=\rho_{n0}=0$ 
is implied by the positivity of the state \cite{Hor85}. Then one
can remove the vacuum-state with a measurement operator $A$ of the form
\begin{align}
A = a_0 \ket{0}\bra{0} + a_1 \sum_{n=1}^\infty \ket{n}\bra{n}.
\end{align} 
with $a_0\ne a_1$:  When $a_0$ is measured one discards the state, for
$a_1$ one continues the experiment and effectively gets the desired
punctured state through post-selection. Recently a
procedure was proposed that realizes the measurement
operator $A$ \cite{Oi13}. 

\subsection{Synthesizing arbitrary states}
The above considerations are only valid for a complete subtraction of
the vacuum state. The procedure does not translate easily 
to more general subtractions, especially for $\alpha_1 \neq 0$,   and
we therefore present another, completely different approach, based on
approximately synthesizing our states from scratch as follows: We can
numerically express the density operator of the desired punctured state
up to a chosen dimension $N$ in
the Fock basis, 
neglecting terms
involving states with higher photon number than $N$:
\begin{align*}
\rho' = \sum_{n=0}^{N}\sum_{m=0}^N c_{mn} \ket{n}\bra{m}.
\end{align*}
This is in general a mixed state and we can express it as a sum of pure states:
\begin{align*}
\rho' = \sum_{i=1} p_i \ket{\psi_i}\bra{\psi_i}\\
\ket{\psi_i} = \sum_{n=0}^N c^{(i)}_n \ket{n}.
\end{align*}
We can then create our state by choosing with the right probability
one of the pure states and constructing it with the method  
proposed in~\cite{Ben93,Law96,Liu04,Hof09a} and demonstrated experimentally in~\cite{Hof09b}. Multiple repetition (creation of an adequate ensemble of
states) then gives us our state up to a certain accuracy. In a single
run we actually do not have the complete state, but for further
experiments one anyway needs typically multiple repetitions for
obtaining measurement statistics of any observable. Hence, this procedure creates the desired mixed state in the ensemble
sense.

\section{Conclusion}

In this work, we introduced a novel class of nonclassical states with
regular non-positive Glauber-Sudarshan $P$ function that we
call punctured states. These states are obtained from the addition of
sufficiently narrow negative peaks to smooth positive
$P$ functions. We determined the regimes of parameters yielding proper
physical (i.e.\ positive semidefinite) nonclassical states in the
case of $\delta$- and Gaussian punctures of thermal or squeezed thermal
states. We showed that their second-order correlation function can be
modified through an appropriate choice of the punctures and 
identify the regimes yielding antibunching of light. All states
exhibiting antibunching have a negative $P$ function in some
region. Finally, we presented possible experimental realizations of
punctured states based on vacuum-or-not measurements and on complete
synthesizing.  

\begin{acknowledgments}
FD acknowledges fruitful discussions with Daniel K.\ L.\ Oi. FD would like to thank the F.R.S.-FNRS for financial support. Computational resources have been provided by the Consortium des \'Equipements de Calcul Intensif (C\'ECI), funded by the F.R.S.-FNRS under Grant No. 2.5020.11.
\end{acknowledgments}

\end{document}